# Driving skyrmions in flow regime in synthetic ferrimagnets

Sougata Mallick[1], Yanis Sassi[1], Nicholas Figueiredo Prestes[1], Sachin Krishnia[1], Fernando Gallego[1], Thibaud Denneulin[2], Sophie Collin[1], Karim Bouzehouane[1], André Thiaville[3], Rafal E. Dunin-Borkowski[2], Vincent Jeudy[3], Albert Fert[1], Nicolas Reyren[1] and Vincent Cros[1]

[1]*Unité Mixte de Physique, CNRS, Thales, Université Paris-Saclay, 91767 Palaiseau, France*
[2]*Forschungszentrum Jülich, ER-C for Microscopy and Spectroscopy with Electrons, Jülich, Germany*
[3]*Laboratoire de Physique des Solides, CNRS, Université Paris-Saclay, 91400 Orsay, France*

**Abstract:**

Despite significant advances in the last decade regarding the room temperature stabilization of skyrmions or their current induced dynamics, the impact of local material inhomogeneities still remains an important issue that impedes to reach the regime of steady state motion of these spin textures. Here, we study the spin-torque driven motion of skyrmions in synthetic ferrimagnetic multilayers with the aim of achieving high mobility and reduced skyrmion Hall effect. We consider Pt|Co|Tb multilayers of various thicknesses with antiferromagnetic coupling between the Co and Tb magnetization. The increase of Tb thickness in the multilayers allows to reduce the total magnetic moment and increases the spin-orbit torques allowing to reach velocities up to 400 m s$^{-1}$ for skyrmions with diameters of about 160 nm. We demonstrate that due to reduced skyrmion Hall effect, combined with the edge repulsion of the magnetic track making the skyrmions moving along the track without any transverse deflection. Further, by comparing the field-induced domain wall motion and current-induced skyrmion motion, we demonstrate that the skyrmions at the largest current densities present all the characteristics of a dynamical flow regime.

**Introduction:**

Magnetic skyrmions, spin swirling particle like entities, have been in the focal point of many researches in condensed matter physics due to their solitonic nature combined with their chiral and topological properties [1–5]. To increase their thermal stability up to room temperature, one of the strategies has been to increase individual layer thickness [6–8] or the number of repetitions in multilayers [9–12]. However, this is at the cost of an enhanced impact of dipolar interaction [13–16], resulting in an enlargement of the skyrmion size and a reduction of the stability for smallest skyrmion [15]. The choice of using multilayers present also an additional challenge as the balance between the various energies at play i.e. exchange, DMI, anisotropy, and dipolar can lead to the stabilization of more complex 3D spin textures with hybrid chirality along the vertical direction [17] or even skyrmionic 3D cocoons [18,19]. A solution to these issues that has been proposed is to rely on materials with reduced or cancelled magnetization i.e., ferrimagnets [7,16] or synthetic antiferromagnets [20–22]. These predictions have been confirmed experimentally [23,24] leading for example to the observation of skyrmions as small as 20 nm at room temperature. Besides the gain in static properties, another advantage of skyrmions in ferrimagnets (or antiferromagnets) is that the so-called skyrmion Hall angle, a transversal motion due to their topological charge, is reduced or even cancelled [25–30].

The combination of a fixed chirality (defined by the sign of interfacial chiral interaction) and their solitonic nature implies that skyrmions can be advantageously displaced through spin-orbit torque (SOT) [3,5,31–33]. However, for most of the skyrmionic systems, an important issue is the presence of a finite pinning landscape [34,35] that impedes them to reach their flow regime of motion. There are predominantly two reasons for this limitation. First, as mentioned before, the impact of skyrmion Hall angle can lead to skyrmion annihilation on the device edges before they can reach very large velocities [36–40]. The second is that the driving force i.e., the amplitude of SOT is not large enough. Hence, in spite of several works reporting high DW velocity in ferrimagnets [24,30,41–45], there are not many reports on the motion of fast skyrmions (>>100 m s$^{-1}$). Recent observations of ferrimagnetic CoGd alloys revealed moderate skyrmion velocity (~150 m s$^{-1}$ at ~ 3 × 10$^{11}$ A m$^{-2}$ however with large diameters of 400-900 nm [37] as well larger ones with 600 m s$^{-1}$ at ~ 1.3 × 10$^{12}$ A m$^{-2}$ for even larger diameters around 1.2 µm [46]).

In this study, we investigate the SOT driven motion of small skyrmions in magnetic-multilayers based on a ferrimagnetic system with the aim of achieving high mobility together with reduced skyrmion Hall effect. To achieve these goals, we consider Pt|Co|Tb multilayers (not the usual choice of CoTb alloys) of various thicknesses with antiferromagnetic coupling between the Co and Tb magnetization, hence forming a synthetic ferrimagnet. In order to increase the velocity of ferrimagnetic skyrmions, there are two important levers that can be activated. The first one is the amplitude of SOT that can be increased by enhancing the charge-to-spin conversion either in the bulk through the choice of nonmagnetic materials in contact with ferromagnet or at the interfaces [47,48]. The second parameter is the energy dissipation rate ($L_\alpha = \alpha^{Co}L_S^{Co} + \alpha^{Tb}L_S^{Tb} = \alpha^{eff}L_S^{eff}$, where α and $L_S = \frac{M_S}{\gamma}$ are the magnetic damping and angular momentum of individual sub-lattices, respectively [16,37,44], that can be efficiently reduced by increasing the antiferromagnetic (AFM) coupling between the Co and Tb layers. Through the tuning of these two parameters, we succeed to reach the flow regime for skyrmion motion at the largest current densities, with velocities up to 400 m s$^{-1}$ for current densities around $8 \times 10^{11}$ A/m². Interestingly, we find that the skyrmions exhibit an almost straight motion in the 1 µm wide tracks. These experimental results demonstrate that our synthetic ferrimagnetic multilayers provide a pinning-free environment for the motion of skyrmions with radius less than 100 nm, highlighting the potential of our system for the development of high-speed and stable skyrmion-based devices.

**Results:**

**Structural characterization:**

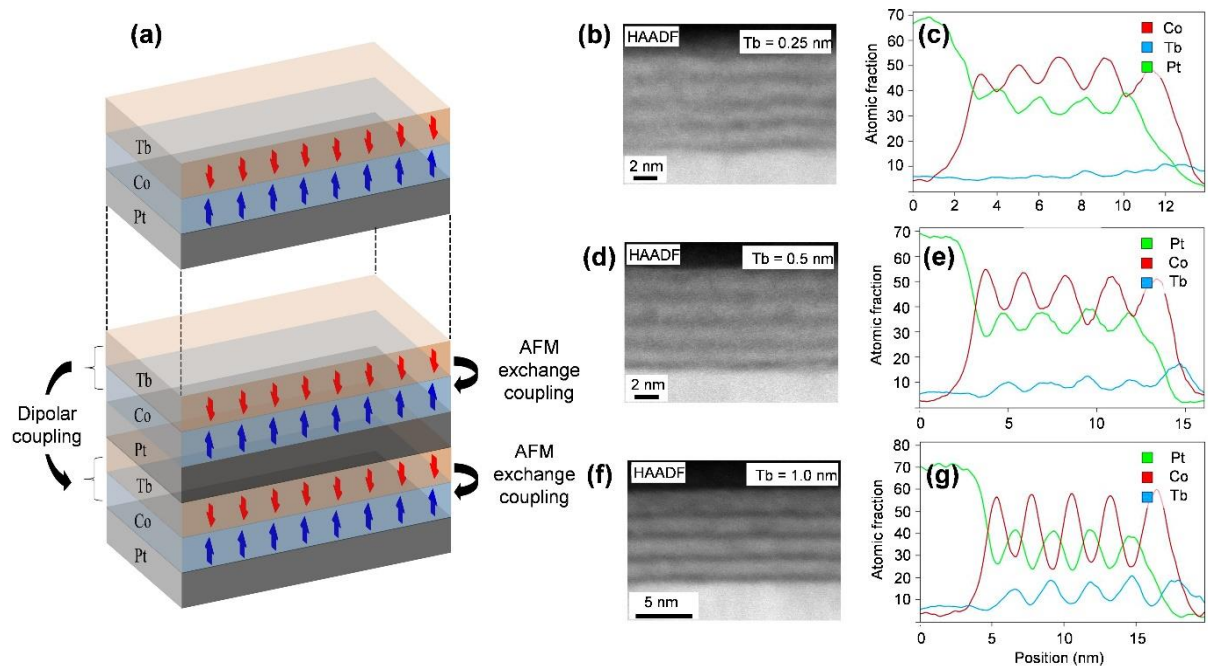

Fig. 1: (a) Schematics of the synthetic ferrimagnetic systems; HAADF STEM images along with the EDX mapping of the thin films (b, c): [Pt (1 nm)/Co(1 nm)/Tb(0.25 nm)]$_{\times 5}$; (d, e): [Pt (1 nm)/Co(1 nm)/Tb(0.5 nm)]$_{\times 5}$; and (f, g): [Pt (1 nm)/Co(1 nm)/Tb(1.0 nm)]$_{\times 5}$.

The sample structure and the coupling between the different constituent layers is schematized in Fig. 1(a), notably with a direct AFM exchange coupling between the Co and Tb layers. Note that there is also an additional interlayer dipolar coupling between different Co-Tb layers inside the multilayer. The Pt layers help in stabilizing the PMA of Co and provides large spin-orbit coupling for generating the interfacial DMI arising in Pt|Co|Tb structure due to breaking of inversion symmetry at both Co interfaces. As reported by S. Alebrand *et al.* [49], the range of composition in which room temperature magnetic compensation can be achieved in CoTb alloy is for Co between 75-80 % and Tb between 25-20 %. Here, we have adopted another strategy in order to mimic the alloy content in the multilayers by choosing the ratio of Co and Tb thicknesses accordingly. The sample quality, and notably the one of the interfaces between Pt, Co, and Tb, has been investigated using scanning transmission electron microscopy (STEM). The HAADF STEM images along with the EDX mapping of the multilayers are displayed in Fig. 1: [Pt (1 nm)/Co(1 nm)/Tb(0.25 nm)]$_{\times 5}$ (Fig. 1 b,c); [Pt (1 nm)/Co(1 nm)/Tb(0.5 nm)]$_{\times 5}$

(Fig. 1 d,e); and [Pt (1 nm)/Co(1 nm)/Tb(1.0 nm)]$_{\times 5}$ (Fig. 1 f,g). In Fig. 1 (b-c), we observe that for $t_{Tb}$ = 0.25 nm (i.e. less than a Tb monolayer), there is no clear peak for Tb in the EDX map (blue curve) and from the HAADF STEM images, meaning that the effect of diffusion and intermixing is significant among the constituent layers for this sample. Beyond the Tb monolayer thickness, for $t_{Tb}$ = 0.5 nm, well-defined layers of individual elements and some partially distinct peaks of Tb are obtained (see Fig. 1 d-e). For $t_{Tb}$ = 1.0 nm (see Fig. 1 f-g), five more distinct peaks of Pt, Co, Tb can be seen in the EDX maps. XRR measurements on the samples optimized for SOT (see discussion later) reveals the presence of a strong intermixing between the Tb and Al at the top interface. However, since the thickness of Tb has been varied continuously from 0.25 – 1.0 nm, we expect to transit from less than a monolayer to continuous layers of Tb albeit with limited accuracy on exact thicknesses.

**Controlling the micromagnetic parameters:**

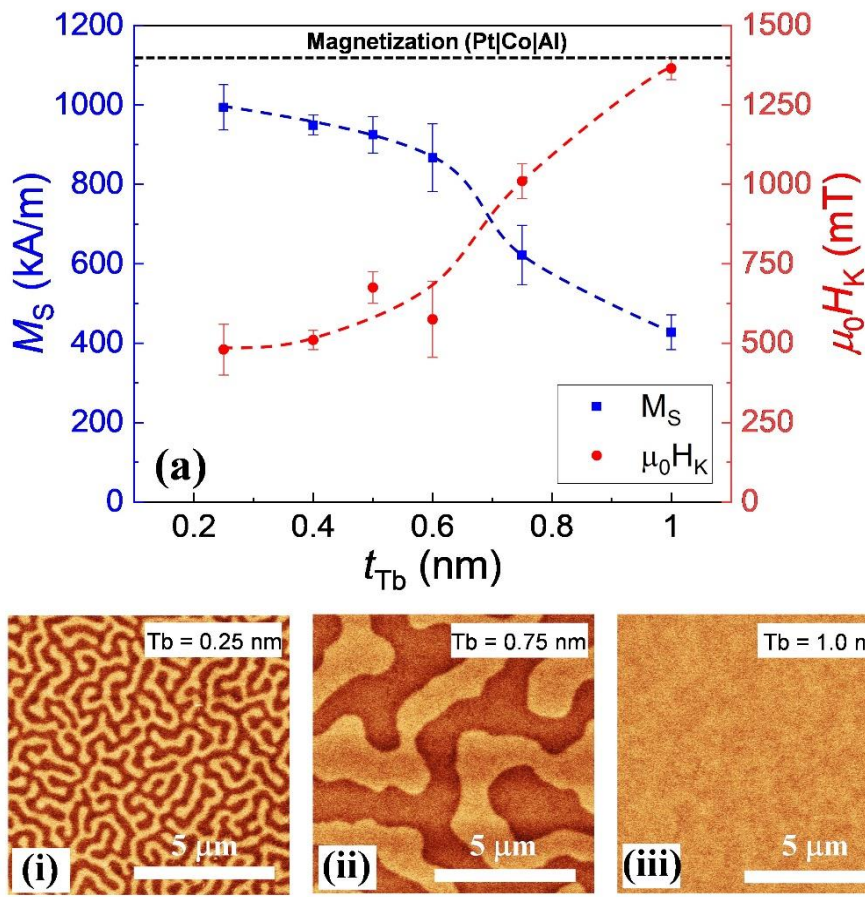

Fig. 2: (a) Magnetization of the [Pt (3 nm)/Co (1.3 nm)/Tb ($t_{Tb}$)/Al (3 nm)]$_{\times 5}$ multilayers as a function of the Tb thickness. The black dotted line represents the magnetization of Pt|Co|Al based multilayers in the absence of any Tb. The blue and red dotted lines are guide to eye for the variation of magnetization and anisotropy, respectively. (b) – (d) Phase signals at demagnetized state measured using conventional MFM for (b: $t_{Tb}$ = 0.25 nm, c: $t_{Tb}$ = 0.75 nm, d: $t_{Tb}$ = 1.0 nm). The color scale bar for all the images is 0.5°.

In Fig. 2 (a), we present the saturation magnetization and the anisotropy field with varying Tb thickness for the //Ta(5 nm)/Pt(5 nm)/[Pt (3 nm)/Co (1.3 nm)/Tb ($t_{Tb}$)/Al (3 nm)]$_{\times 5}$ multilayers. Note that Pt thickness inside the multilayers has been increased from 1 to 3 nm compared to previous series in order to enhance the PMA of Co. We find a decrease in magnetization and an enhancement of PMA with increasing Tb thickness, both being associated to the enhancement of the AFM coupling between the Co and Tb layers. This increase of coupling between Co and Tb films has been confirmed by XMCD measurements both at Co and Tb edges. The effective anisotropy fields extracted from in-plane magnetometry are 480 ± 80 mT, 510 ± 30 mT, 575 ± 120 mT, 1010 ± 55 mT and 1365 ± 35 mT for $t_{Tb}$ = 0.25, 0.4, 0.6, 0.75, and 1 nm, respectively (see Fig. 2 a). In the Supplementary Information (Fig.

S2), we provide similar measurements for 1 nm Pt series, in which we also reveal that the magnetization decreases monotonically with the increase of Tb thickness from about an atomic layer (~ 0.3 nm) up to about 3 atomic layers (1 nm). The difference in Pt 1 nm series, is however that for the sample with thickest Tb layer, i.e., $t_{Tb} = 1$ nm, the anisotropy switches from perpendicular to in-plane.

Another important parameter in skyrmion multilayers is the amplitude of the chiral exchange interaction i.e., the interfacial DMI. To determine it properly, BLS measurements in Damon-Eschbach geometry have been performed (see supplementary figure S1 for discussion). We find $D_s = -1.62 \pm 0.20$ pJ/m for the single layer ($N = 1$) sample with //Ta(5 nm)/Pt(8 nm)/Co(1 nm)/Tb(0.25 nm)/Al(3 nm). This $D_s$ valure is found to agree very well with the expectation that iDMI is related to the work function difference of the material on top of Co film [50]. The presence of such large interfacial DMI together with the interlayer dipolar coupling results in the stabilization of skyrmions in the multilayer samples up to $t_{Tb} = 0.6$ nm. However, for thicker Tb (up 1 nm), the anisotropy energy arising from the AFM coupling between Co and Tb becomes too large, impeding the formation of any skyrmionic phase (see Fig. S4 in Supplementary Information). In Fig. 2 (i) – (iii), we present the MFM phase signal of the demagnetized states for the multilayer samples [Pt (3 nm)/Co (1.3 nm)/Tb ($t_{Tb}$)/Al (3 nm)]$_{\times 5}$ with $t_{Tb} =$ 0.25, 0.75 and 1 nm respectively. In $t_{Tb} = 0.25$ nm sample, we observe the formation of typical labyrinthic domains of PMA systems. In $t_{Tb} = 0.75$ nm sample, the domain width and the periodicity increase substantially as expected from the PMA enhancement. Finally, for $t_{Tb} = 1$ nm, the magnetic anisotropy is so large that the sample cannot be demagnetized by this procedure.

**Field-induced DW motion:**

In order to explore different mobility regimes, and notably to extract the parameters describing the pinning potentials, we have investigated the field-induced motion of domain walls (DW) using magnetic field pulses applied in the out-of-plane direction. The DW velocity as a function of the external field for the sample //Ta(5 nm)/Pt(8 nm)/Co(1 nm)/Tb(0.5 nm)/Al(3 nm)/Pt(2 nm) detected through Kerr microscopy is shown as black stars in Fig. 3. The obtained results can be fitted with the self-consistent description of the creep (dashed red line) and depinning (solid green line) transition with the following equations [51,52]:

$$v = v(H_d, T) e^{-\frac{\Delta E}{kT}} \text{ with } \Delta E = k_B T_d \left( \left[ \frac{H}{H_d} \right]^{-\mu} - 1 \right) : \text{creep regime } (H \leq H_d)$$

$$v(H_d, T) = v_T(H_d, T) \cdot \left( \frac{T}{T_d} \right)^{\psi} : \text{depinning regime } (H = H_d)$$

$$v(H, T) = \frac{v_T(H_d, T)}{x_0} \left( \frac{H - H_d}{H_d} \right)^{\beta} : \text{depinning regime } (H > H_d) \text{ in the a-thermal limit}$$

where $\mu = 0.25$, $\beta = 0.25$, $\psi = 0.15$, and $x_0 = 0.65$ are the universal parameters for thin films with short range pinning, whereas $H_d, v_T, T_d$ (depinning field, velocity, and temperature) are the non-universal fitting parameters [53,54]. In the inset of Fig. 3, we demonstrate the transition between the creep and depinning regimes. We note that the depinning temperature extracted from the fitting is ~ 8000 K which is similar to the values obtained for Pt/Co/Pt systems [55].

Furthermore, from the fitting parameters in both creep and depinning regime, we can predict the DW flow regime (dotted wine line in Fig. 4) using the following equation:

$$v = v_T(H_d, T) \frac{H}{H_d}$$

Interestingly, the current driven skyrmion motion (shown as blue stars in Fig. 3) can be superimposed to the field-induced DW motion using a conversion factor: $\frac{\mu_0 H}{J} = 27.5 \times 10^{-14}$ Tm$^2$A$^{-1}$. First, it can be seen that there is a similar behavior between DW and skyrmions in the creep and in the depinning regime for these two kinds of driving mechanisms. Indeed, the two trends depart from each other for the largest current densities i.e., the largest equivalent field, for which we find that only skyrmions are reaching velocities in agreement with predictions for a flow regime i.e. a linear increase of velocity with driving force (see dotted line in Fig. 3). To our knowledge, this result represents the first observation of a 'real' steady state current-induced skyrmion motion.

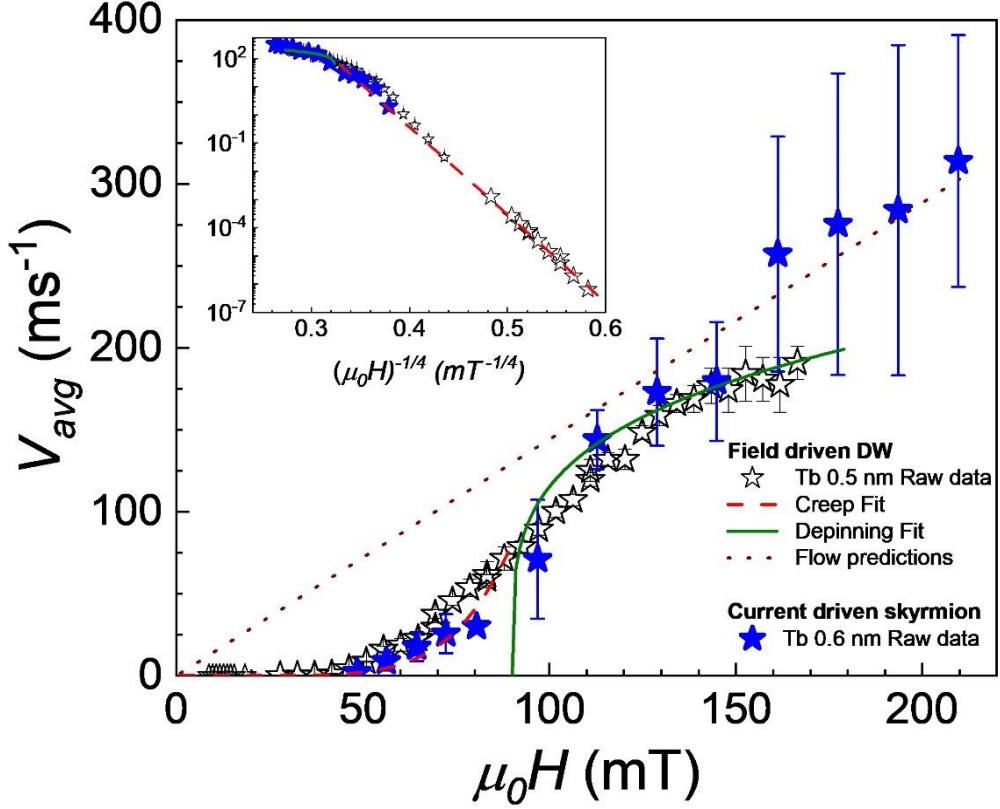

Fig. 3: Comparison between field-induced DW motion (open black stars) and current induced skyrmion motion (solid blue stars) with prediction of DW creep (dashed red line), depinning (solid green line), and flow (dotted wine line). Inset: same curves plotted in semi-log scale versus $(\mu_0 H)^{-1/4}$, evidencing the creep regime.

**Optimization of SOT for skyrmion dynamics**

Second harmonic Hall measurement technique has been used to quantify the damping-like (DL) torque. The amplitude of DL torque ($\mu_0 H_{DL}$) are $(0.40 \pm 0.02) \times 10^{-11}$ and $(0.57 \pm 0.03) \times 10^{-11}$ mT A$^{-1}$ m$^2$ for $t_{Tb} = 0.25$ nm and 0.5 nm respectively, in //Ta(5 nm)/Pt(8 nm)/Co(1.0 nm)/Tb($t_{Tb}$)/Pt(2 nm). Subsequently, the effective spin-Hall angle of each sample has been estimated by using:

$$\theta_{SHE}^{eff} = \frac{2e\mu_0 H_{DL} M_S t}{\hbar J_{Pt}}$$

The effective spin-Hall angles are found to be 1.9 % and 3.4 %, respectively, for the samples with these two samples $t_{Tb} = 0.25$ nm and 0.5 nm. The total effective spin-Hall angle is hence rather small (compared to the one of Pt ~ 9 % [56]), which is mainly due to the partial cancellation of spin Hall effect arising from top and bottom Pt layers. Recently, we have demonstrated that inserting a 3 nm Al layer in between the Co and top Pt allows to strongly decrease the transmission coefficient at the top Al/Pt interface, resulting in an increase of the effective spin Hall angle and the related DL torque up to 30 % [56]. Following the same strategy between the Tb and top Pt layers, we have inserted a 3 nm thick Al layer in our heterostructure //Ta(5 nm)/Pt(8 nm)/Co(1.0 nm)/Tb($t_{Tb}$)/Al(3 nm)/Pt(2 nm). As expected, we find an enhancement of DL torques $\mu_0 H_{DL} = (2.14 \pm 0.02) \times 10^{-11}$ and $(1.99 \pm 0.08) \times 10^{-11}$ mT A$^{-1}$ m$^2$ for $t_{Tb} = 0.25$ nm and 0.5 nm, respectively. In these samples including Al layers, the calculated effective spin-Hall angles are 8.1 % and 8.6 %, respectively for the samples with $t_{Tb} = 0.25$ nm and 0.5 nm.

**Current-induced motion of skyrmions:**

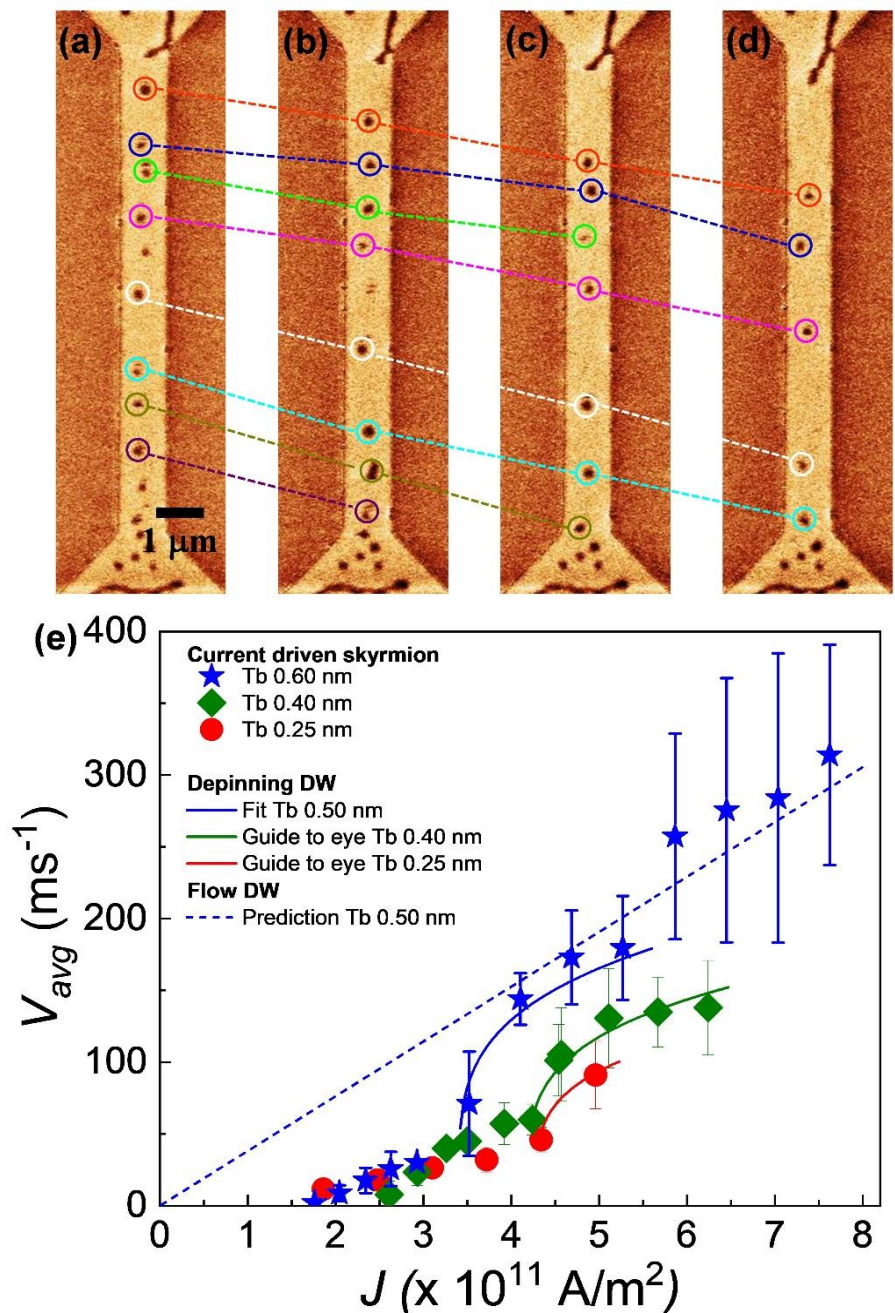

Fig. 4: (a) – (d) Current induced skyrmion motion in [Pt (3 nm)/Co (1.3 nm)/Tb (0.6 nm)/Al (3 nm)]$_{\times 5}$ multilayers after application of two successive 10 ns wide current pulses in between two images with $J = 3.5 \times 10^{11}$ A/m$^2$ under an external perpendicular field ~41 mT. The circles and the dashed lines are guide to eye to track the motion of individual skyrmions. (e) Average skyrmion velocity (filled symbols) vs current density for different samples with varying Tb thickness in the multilayers. Error bars corresponding to each data points represent the standard deviation of the skyrmion velocity inside the tracks. The blue straight and dashed lines are reported from Fig. 3, and correspond to the fit of depinning law and the prediction for the flow regime, respectively. The green and red lines are guides to eye underlying the depinning law.

In the following discussion, we focus on the [Pt (3 nm)/Co (1.3 nm)/Tb (0.6 nm)/Al (3 nm)]$_{\times 5}$ multilayer. Note however that the general behaviour that will be described has been found similar for the multilayers with $t_{Tb} = 0.25$ and 0.4 nm. In the sample with Tb = 0.6 nm, the apparent diameter of skyrmions is found to be between $160 - 190$ nm (see Fig. 4(a)). The current driven motion of skyrmions under an external out-of-plane field of ~ 41 mT is illustrated by a sequence of MFM images in Fig. 4(a) – (d). Each frame has been acquired after application of two successive 10 ns long current pulses (from top to bottom in the figures) with $J = 3.5 \times 10^{11}$ A/m$^2$. Note that the current densities have been calculated by considering a uniform current distribution flowing through the whole thickness of the samples. The colored open circles and the dashed lines in Fig. 4 are guides to the eye to follow the motion of individual skyrmions inside the track. Movies of skyrmion motion are presented as supplementary video files. First, we observe that the position of the skyrmions after a series of pulses remain in the central part of the 1 µm wide track, with no clear transverse displacement due to skyrmion Hall effect. This relative straight motion is due to the combination of the reduced skyrmion Hall angle expected in ferrimagnet and the finite repulsion from the track edges. This explanation is further confirmed by the observation of a transverse skyrmion motion in a wider track (3 µm), in which skyrmions are found to move with an angle of ~29°. This corresponds to a non-complete cancellation of the transverse displacement but largely reduced compared to the ones observed for skyrmions in ferromagnetic multilayers [25–27,31,36,57] (see supplementary Fig. S3 for comparison). The experimental skyrmion Hall angle can be compared to the one estimated using magnetic parameters extracted from expected $\theta_{SHA} = \arctan\left(\frac{1}{\alpha}\frac{2\Delta}{R}\right)$ with α, Δ, and $R$ being the magnetic damping, the domain wall width and the skyrmion radius respectively. The damping parameter estimated from BLS measurements is found to be α = 0.13, 0.25, and 0.43 for Tb = 0.25, 0.5 and 1 nm respectively. Considering an exchange stiffness equal to $A = 10$ pJ/m [50,58], the calculated skyrmion Hall angles are 45°, and 27.5° for the samples with $t_{Tb} = 0.25$, and 0.5 nm respectively. Thus, we conclude that the experimentally obtained value of ~29° is in good agreement to the predicted value of 27.5°. As expected from the expression of the skyrmion Hall angle, we confirm that the reduction of the skyrmion Hall effect is directly related to the increment of the effective magnetic damping which is a consequence of the antiferromagnetic coupling between the Co and Tb layers [59].

In Fig. 4(e), we display the experimental skyrmion velocity *vs.* current density for different samples with varying Tb thickness, namely 0.25 nm, 0.4 nm and 0.6 nm (see Supplementary Fig. S4 for current dynamics experiments for samples with thicker Tb). Each velocity is obtained by the measuring the displacements for all the skyrmions (typically between 5 and 10) present in the 10-20 µm long track. The largest input current density is the one for which no additional skyrmion nucleation is occurring. The error bars correspond to the standard deviation of these velocities. Note that the rather large distributions of velocity indicate that the motion of all the skyrmions inside the track is not exactly the same after each current pulse, which further hints towards presence of scarce pinning sites within the samples. More importantly, at higher current densities, there is a lesser number of skyrmions present in the track due to the large distance of propagation in between successive pulses (larger than 2 µm in average). As a consequence, only a few events inside the track can be recorded for each current pulse values, hence reducing the statistics.

From the comparison between DW and skyrmion motion discussed previously (see Fig. 3), different dynamical regimes can be clearly identified in Fig.4(e). The positive curvature of the velocity curves is observed systematically, strongly suggesting that the skyrmions present a depinning transition. The good agreement with the prediction for DW depinning law in the a-thermal limit (shown in Fig. 3) suggests that skyrmions could also present universal behaviours with a power law variation of the velocity with the current density. We notice that the depinning threshold is found to increase slightly with increasing Tb thicknesses, which may result from an enhancement of the skyrmion-disorder interaction (whose discussion is well beyond the scope of this paper). Below the depinning threshold, the motion of the magnetic textures is expected to be thermally activated. Therefore, the quasi-linear variation of the velocity observed at low current densities in Fig. 4(e) corresponds to the thermally activated motion of skyrmions. Note however that their velocity does not follow the scaling law (not shown) ($lnv \sim j^{-1/4}$) characteristic of DW creep motion [52,54]. This different behaviour is expected since DW elasticity plays a major role for DW creep motion, while small skyrmions should be more rigid. Above the depinning transition, the skyrmion motion is predicted to reach the flow regime which is independent of pinning and controlled by dissipation. As observed in Fig. 4(e), the flow regime is only reached for $t_{Tb} = 0.6$ nm (see also Fig. 3). For the other multilayers, this regime is in fact not reached mainly because it is impeded by multiple nucleation of skyrmions at large current density. Note that the maximum velocity ~ 400 m/s (specifically for skyrmions with radius < 100 nm) is among the largest ones for dynamics of ferrimagnetic skyrmions [37,46].

Although the effective spin Hall angle increases from 8.1 % to 8.6 % in the thickness range of $t_{Tb}$ = 0.25 to 0.5 nm, this cannot exclusively elucidate the strong increase in the skyrmion velocity between the two samples. The origin of such large skyrmion velocity can be further explained following the approach of L. Berges *et al.* [37]

$$|v| = \frac{v_0}{\sqrt{1+\rho^2}}$$

where $\rho$ is the rate of skyrmion deflection (skyrmion Hall angle is $\arctan(\rho)$) and $v_0$ is the velocity of skyrmion without any deflection ($v_0 = \frac{\hbar J \theta_{SHE}}{2eL_\alpha t} \frac{f}{d}$, predominantly depends on the SOTs and energy dissipation). As mentioned before, the skyrmion Hall angle is reduced when the Tb layer thickness is increased due to increasing damping as $\rho = \frac{n}{\alpha d}$ where $n$, $d$, and $f$ correspond to texture topology, magnetization rotation length scale, and texture chirality, respectively [37]. The ratio $\frac{n}{d} = \frac{2\Delta}{R}$ is a constant when the skyrmion radius ($R$ ~ 80 - 95 nm) is significantly larger than the domain wall width parameter ($\Delta = \sqrt{\frac{A}{K_{eff}}}$ ~ 6 nm). Hence, increasing Tb thickness leads to decrease in $\rho$, which in turn results in an increased skyrmion velocity $v$. Additionally, increasing the Tb thickness leads to the enhancement of the Gilbert damping parameter $\gamma$ (data not shown here), which together with reduced $M_S$ leads to lowering of $L_S$ ($= \frac{M_S}{\gamma}$) and consequently the energy dissipation. Therefore, these results confirm that the enhanced skyrmion velocity in our system is a consequence of the systematic engineering of the efficient damping-like torque as well as the AFM coupling between the Co-Tb layers.

**Discussion:**

We have demonstrated the flow regime of skyrmions induced by SO torques in tracks made of synthetic ferrimagnets. Through a fine tuning of the properties of Pt|Co|Tb multilayers, notably the strength of the antiferromagnetic coupling between Co and Tb films and the amplitude of the SO torques, we succeed to increase the skyrmion velocity in the viscous regime i.e. immune to pinning centers, up to 400 m/s for skyrmions with diameters of about ~160 nm. Additionally, the reduced skyrmion Hall effect and edge repulsion of the magnetic track allow the skyrmions to move along the 1 μm narrow track without any transverse deflection. For the optimum Tb thickness of 0.6 nm, the current-induced skyrmion dynamics is found to be quantitatively equivalent to the field-driven domain wall ones both in the creep and depinning regimes before they differ at higher current densities in which skyrmions are showing significantly better velocities. Such improved mobility originates from the optimization of the SO torque efficiency, notably through the insertion of a thin Al layer in the multilayer between Tb and Pt films in order to better control the spin current transmission coefficient and hence the SOT amplitude. Our findings in synthetic ferrimagnetic systems presents a straightforward method of tuning the antiferromagnetic coupling, the micromagnetic energies, and the SOT amplitude, leading to skyrmion dynamics going beyond what is achieved in more standard ferromagnetic multilayers. We believe that these results might pave the way towards the development of low energy and fast skyrmion based spintronic devices.

## Methods:

### Sample preparation and characterization:

Samples have been grown at room temperature in a high-vacuum sputtering chamber with a base pressure of $5 \times 10^{-8}$ mbar on thermally oxidized Si substrates. The sputtering offers an industry compatible process with excellent reproducibility with a precision of individual layer thickness up to 0.03 nm. All the multilayers are grown on top of Ta (5 nm)/Pt (8 nm) buffers to help stabilizing PMA in Co and capped with 3 nm Pt to protect from oxidation. The multi-layered samples have the following structure: //Ta (5 nm)/Pt (7 nm)/[Pt (1 nm)/Co ($t_{Co}$)/Tb ($t_{Tb}$)]$_{\times N}$/Al (2 nm), where $t_{Co} = 1.0 - 1.5$ nm, $t_{Tb} = 0.25 - 1.0$ nm, and $N = 1$ and 5, // indicates the thermally oxidized Si substrate, covered by 280 nm of $SiO_2$.

Thickness and roughness calibration of the layers have been performed using x-ray reflectivity (XRR) measurements. The in-plane and out-of-plane magnetization measurements have been performed using an alternating gradient field magnetometry (AGFM) setup. The quality of the interfaces in the samples, between Pt, Co, and Tb, has been investigated using scanning transmission electron microscopy (STEM). DMI in the heterostructure has been extracted by performing Brillouin light scattering (BLS) experiments in Damon-Eschbach geometry. The effective contribution of field-like and damping-like torques in the samples have been measured by performing 2nd harmonic Hall measurements in a PPMS set-up. Field induced dynamics of the DWs in the samples have been performed using a Kerr microscopy set-up where the external field pulses are applied using a micro-coil placed below the samples. Total thickness of combined Co and Tb layers in the samples has been considered for the calculation of various micromagnetic parameters.

### Magnetic imaging:

The observation of the magnetic skyrmions have been performed using a magnetic force microscopy (MFM) set-up at room pressure and temperature. Dynamic magnetic imaging of the skyrmions was performed on a stage customized for high frequency transport measurements. The MFM tips are home made with a thin CoFeB (5 nm) layer coated by Al layer to reduce magnetic perturbation during topography scans (more discussions are provided in the supplementary information). In order to eliminate heating and thermal drifts, permanent magnets have been used to apply the external magnetic field during the measurements.

In order to study the SOT driven skyrmion dynamics in tracks patterned in our synthetic ferrimagnetic structure, before each set of experiments, skyrmions are first nucleated from a uniform magnetic state under a suitable external perpendicular field in the range 30-45 mT by applying a 20 ns current pulse with $J > 3 \times 10^{11}$ A/m$^2$. Subsequently, the skyrmions are driven with shorter (10 ns) pulses to reduce additional parasitic nucleation due to Joule heating.

### Author contributions: